\documentclass{ws-p8-50x6-00}
\newcommand{\mum}{$\,\mu$m}

\begin{document}

\title{The brighter side of sub-mm source counts: 
       a SCUBA scan-map of the Hubble Deep Field}

\author{Colin Borys$^\dag$, Scott Chapman$^\ddag$,
        Mark Halpern$^\dag$,Douglas Scott$^\dag$}
\address{$^\dag$Department of Physics \& Astronomy,
         University of British Columbia, 
         Vancouver BC CANADA V6T 1Z1\\E-mail: borys@physics.ubc.ca}

\address{$^\ddag$Observatories of the Carnegie Institution of Washington,
         Pasadena CA USA 91101}

\maketitle

\abstracts{ 
We present an $11\times11$ arcminute map centred on the Hubble Deep 
Field taken at 850\mum\ with the SCUBA camera on the JCMT.  The map has 
an average one-sigma sensitivity to point sources of about 2.3 mJy and 
thus probes the brighter end of the sub-mm source counts. We find 7 
sources with a flux greater than 9\,mJy ($\sim 4\sigma$), and therefore 
estimate $N(>9\,\rm{mJy})= 208^{+90}_{-72}\,\rm{degree}^{-1}$.  This 
result is consistent with work from other groups, but improves the
statistics at the bright end, and is suggestive of a steepening of the
counts.
}

\section{Introduction}
Observations using the Submillimetre Common User Bolometer Array 
(SCUBA\cite{scuba}) have been revolutionizing our understanding of the 
importance of dust in galaxies at high redshift.  Already several very 
deep integrations have been carried out on single SCUBA fields down
to the confusion limit.  In order to learn more about source counts 
(and hence to constrain models), the next step is to search for brighter
objects over somewhat larger fields.

The population of bright sub-mm sources is currently not well understood.
Current models\cite{blain,eales,blainlong,fall} for source counts in 
the sub-mm have
been able to account for the observed sources by invoking evolution 
which follows the $(1+z)^3$ form required to account for IRAS galaxies 
at 60$\,\mu$m, and the powerful radio-galaxies and quasars\cite{dunlop}.
Euclidean models with no evolution have a 
slope of roughly $-1.5$ ($N_s \propto S_{\nu}^{-3/2}$), which cannot 
possibly account for the sources observed. With reasonable evolution 
(in IRAS-motivated models), the counts steepen sharply at the 10's of 
mJy sensitivity level to roughly $S_{\nu}^{-2.6}$.  At the source 
detection level of previous work, typically 5\,mJy, there is little 
variation between the models, and we are well into the steep counts 
regime.

However, at the $10$--$30\,$mJy sensitivity level various evolutionary 
models\cite{guid} show more parameter dependence.  Given a possibly wide 
range of galaxy types contributing to the source counts at the bright 
end, the actual counts may deviate from a simple parametric model 
dramatically.  Furthermore, cosmological, as well as evolutionary 
parameters, play a role here.

\section{Data collection and analysis}
The standard scan-map observing strategy is to 
use multiple chop throws in two fixed directions on the sky 
so that an Emerson deconvolution technique can be applied.  This may
be appropriate for galactic plane mapping, where structure appears
on all scales, but we seek a simpler approach which is optimized for 
finding point sources. The  disadvantage of the standard approach is 
that the off-beam pattern gets diluted (making it more challenging 
to isolate faint sources), and 
the map noise properties are more difficult to understand. 
Our approach was to use a single chop direction and throw
fixed on the sky. The result is a map that has, for each source, a 
positive and negative signal. 

A total of 61 scans were obtained, 31 in early 1998 and 30 more a year 
later.  The data were first analysed using SURF, but because the data 
were taken in a non-standard mode, we found it necessary to write custom 
software.  In this way, we were able to isolate and remove large scale 
features in the maps, and estimate the per pixel noise level by a 
careful accounting of the per-bolo noise and the frequency at which it 
sampled a particular pixel. 

Sources were found by using a model for the dual beam pattern and 
fitting it, in a least-squares sense, to each pixel in the map. A total 
of 7 sources were found that had a peak to error-of-fit ratio 
greater than
4.  We also rotated the dual beam pattern by ninety degrees and used 
that as a model; Only 2 sources were found, one of which is associated 
with the brightest source in the map.  Monte-carlo simulations suggest 
that the other false positive is not unexpected.  As an additional 
check, the 1998 data were compared with the 1999 data to ensure that each
source was evident in both halves of the data.

The calibration was determined by fitting the beam model to observations
of standard calibrators taken during the run.  We checked our procedure 
by re-analysing the Barger\cite{barger_radio} et al. HDF/radio fields, 
which are a set of jiggle maps within the HDF flanking fields.  In fact 
this led to a correction in our original flux estimates.  Figure 1 
presents our final map, and an updated source count plot is given in 
Figure 2.

\begin{figure}[ht]
\epsfxsize=18pc
\center\epsfbox{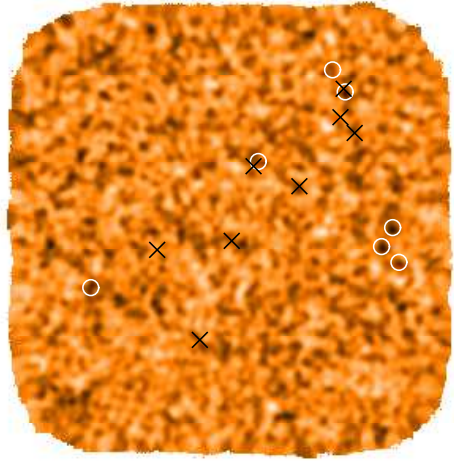} 
\caption{The 850\mum\ HDF map. \label{fig:hdf850}
The map size is roughly $12 \times 12$ arcminutes, although 
only the  central $11 \times 11$ was used in order to avoid the noisy 
edges.  The chop is 40 arcseconds and is roughly east-west.
The circles outline the 7 sources detected in our survey and the
dark crosses highlight sources found in the Barger et al.\,(2000) and 
Hughes et al.\,(1998) work.  Since these surveys went deeper, not all 
can be detected in the scan map. They also cover a smaller area, and
hence cannot see several of our sources.}
\end{figure}

\begin{figure}[ht]
\epsfxsize=20pc 
\center\epsfbox{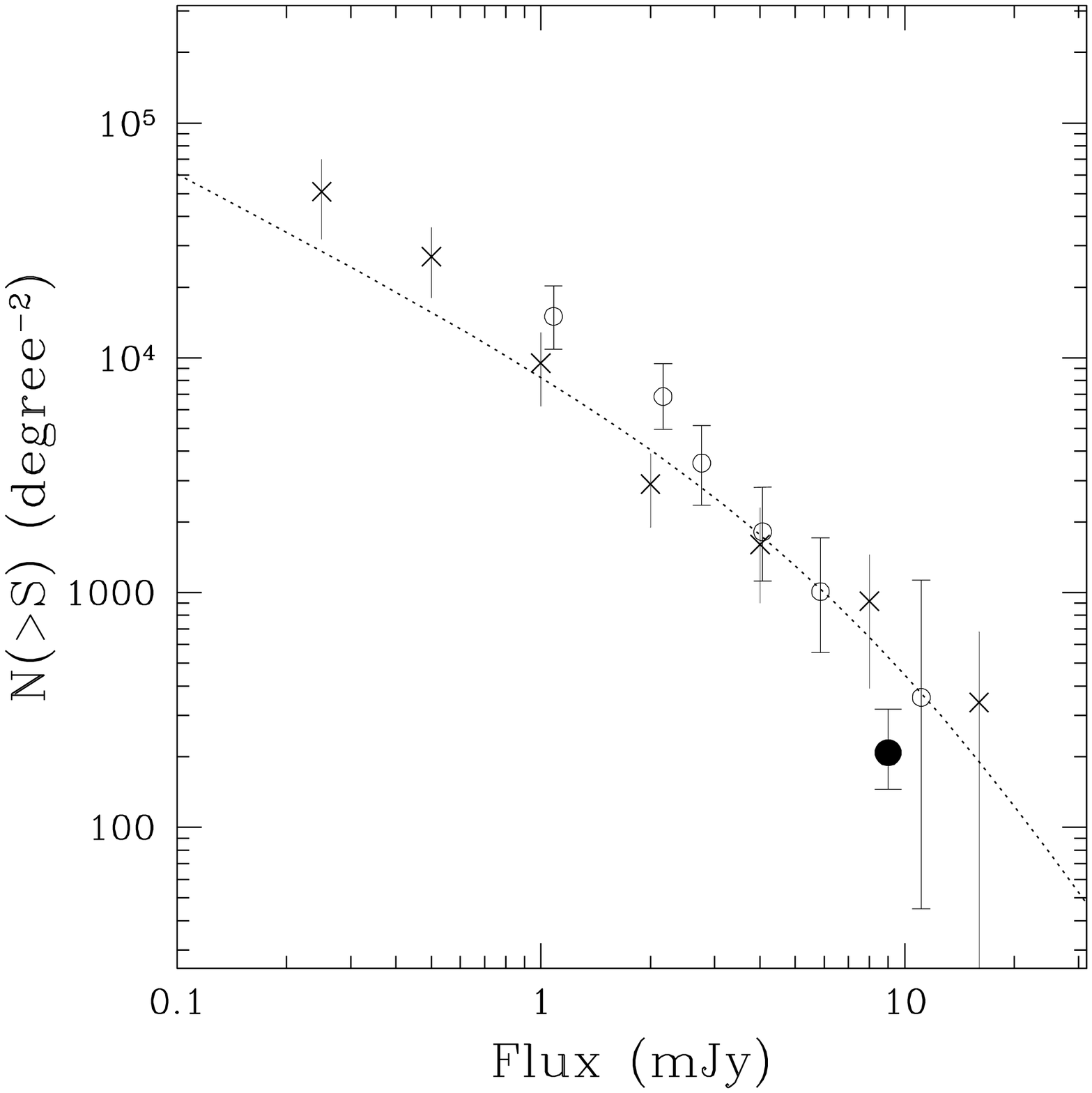} 
\label{fig:counts850}
\caption{The 850\mum\ source counts. 
The crosses are data from Blain et al.\,(1999) and the open circles 
that from a
recent cluster survey by Chapman et al.\,(2000)  The dashed line
is a simple two power-law fit to the previous counts. Our new 
estimate, with Bayesian 68\% confidence limits, is given by the 
solid circle. 
}
\end{figure}

\section{Conclusions and followup work}
We have significantly improved statistics at the $\sim\,10\,$mJy level, 
and suggest there is some indication of a steepening of the counts 
there.  Because these sources are brighter than typical SCUBA detections,
they will be relatively easy to follow up at other wavelengths.  
Preliminary analysis already indicates good correlation with $\mu$Jy 
radio sources, but little indication of optical counterparts 
(as found in other studies).


Although not discussed here, we are also investigating sources in the
450\mum\ scanmap.  However, this is more arduous given the lower 
sensitivity of SCUBA with this filter.  Finally, a similar analysis is 
being conducted on data taken of 50 square arminutes of the 
Groth strip/Westphal region.

\section*{Acknowledgments}
We would like to thank Amy Barger for access to her data before it 
was available on the JCMT public archive, and Doug Johnstone for 
several useful conversations.  We are also grateful to the staff at the 
JCMT, particularly Tim Jenness and Wayne Holland who provided invaluable
advice.


\begin{thebibliography}{99}

\bibitem{scuba} Holland, W.S., et al., \Journal{MNRAS}{303}{659}{1999}



\bibitem{blain} Blain, A.W., \Journal{MNRAS}{295}{92}{1998}

\bibitem{eales} Eales S., Edmunds M.G.,
\Journal{MNRAS}{280}{1167}{1996}

\bibitem{blainlong} Blain A.W., Longair M.S., 
\Journal{MNRAS}{279}{847}{1996}

\bibitem{fall} Fall S.M., Charlot S., Pei Y.C.,
\Journal{ApJ}{464}{L43}{1996}

\bibitem{dunlop} Dunlop J.S., Peacock J.A.,
\Journal{MNRAS}{247}{19}{1990}

\bibitem{guid} Guiderdoni B., et al., \Journal{MNRAS}{295}{877}{1998}


\bibitem{barger_radio} Barger, A. J.; Cowie, L. L.; Richards, E. A.,
\Journal{AJ}{119}{2092}{2000}


\bibitem{hughes} Hughes D.H., et al., \Journal{Nature}{394}{241}{1998}
\bibitem{blain2} Blain A.W., et al., 
\Journal{ASP conference proceedings}{193}{246}{1999}

\bibitem{chapman} Chapman, S.C., et al., 
\Journal{MNRAS}{submitted}{astro-ph/0009067}{2000}.

\end{thebibliography}
\end{document}